\documentclass[12pt]{article}

\newcommand{\x}{arXiv:}
\newcommand{\m}{\mathrm}
\newcommand{\be}{\begin{equation}}
\newcommand{\ee}{\end{equation}}
\newcommand{\ba}{\begin{eqnarray}}
\newcommand{\ea}{\end{eqnarray}}

\usepackage{graphicx}
\usepackage{amssymb}
\usepackage{amsmath}
\usepackage[T1]{fontenc} 
\usepackage[ansinew]{inputenc} 
\usepackage[nosort]{cite}
\newcommand{\inbar}{\vrule height1.57ex width.4pt depth0pt}
\newcommand{\SW}{\relax{\hbox{$\ \inbar\kern-.285em{\rm S}$}}}

\begin{document}
\thispagestyle{empty}
\begin{center}

\null \vskip-1truecm \vskip2truecm

{\Large{\bf \textsf{Jet Quenching in The Most Vortical Fluid:}}}

{\Large{\bf \textsf{A Holographic Approach}}}

{\large{\bf \textsf{}}}

\vskip1truecm

{\large \textsf{Brett McInnes
}}

\vskip0.1truecm

\textsf{\\ National
  University of Singapore}
  \vskip1.2truecm
\textsf{email: matmcinn@nus.edu.sg}\\

\end{center}
\vskip1truecm \centerline{\textsf{ABSTRACT}} \baselineskip=15pt
\medskip

The STAR collaboration at the RHIC facility has recently announced the exciting discovery of direct evidence for extremely large vorticity in the Quark-Gluon Plasma generated in peripheral collisions, seen in the form of global polarization of $\Lambda$ and $\bar{\Lambda}$ hyperons. This prompts the question: does this vorticity have any effect on other observed phenomena, such as jet quenching? Using a simple gauge-gravity model, we suggest that such an effect may be detectable, in data from near-future experiments, as a reduction in the jet quenching parameter. This is predicted to be most prominent in collisions corresponding to a narrow range of centralities around $10 \%$. The relative reduction (compared to collisions at zero or, alternatively, large centrality) is predicted to behave in an unexpected and characteristic manner: the model suggests that it is independent of the impact energy.

\newpage
\addtocounter{section}{1}
\section* {\large{\textsf{1. The Most Vortical Fluid}}}
It has long been predicted \cite{kn:liang,kn:bec,kn:huang,kn:viscous,kn:deng,kn:vortical,kn:hyper} that peripheral heavy-ion collisions, as studied in the RHIC and LHC facilities, will generically give rise to a Quark-Gluon Plasma (QGP) with a \emph{very large vorticity}. This vorticity should manifest itself in the form of global polarization of $\Lambda$ and $\bar{\Lambda}$ hyperons. In practice, however, the technical difficulties of such observations are formidable, and past searches for similar phenomena \cite{kn:abel} were unsuccessful. It was therefore a major advance when the STAR collaboration \cite{kn:STAR} at the RHIC facility announced \cite{kn:STARcoll} the first direct experimental data in which this effect plays a clear role. (See \cite{kn:lambdahui,kn:lambdabec,kn:lambdahannah,kn:vorticityoverview} for discussions of the significance of this discovery.)

The reported values of the vorticity (around $9\,\pm 1\,\times 10^{21}\,\cdot\,$s$^{-1}$, $\sqrt{s_{NN}}$-averaged \cite{kn:STARcoll}) are such as to prove that the QGP in this case is the ``most vortical'' fluid ever observed. In fact, the vorticity field produced in individual collisions is a complicated function of many variables, such as position \cite{kn:lambdahui}, time \cite{kn:yifeng}, and of course the impact parameter; so from a theoretical perspective one should be prepared to associate a wide range of local angular momentum densities with this overall value. In short, the actual situation in some cases may be still more extreme than the reported figure suggests.

One naturally wonders whether such strong vorticity fields can affect other important QGP-related phenomena currently being studied. For example, one of the most important ways to understand the QGP is through the study of \emph{jet quenching} \cite{kn:jet,kn:jetagain}, represented by $\hat{q}$, the mean squared transverse momentum accumulated (per distance travelled) by a hard parton traversing the plasma. Is it possible that the vorticity is so large that (in some cases) it can affect $\hat{q}$? This is the question we will attempt to address in this work.

There are in fact hints that effects of this kind can be observed: for example, it has been pointed out (see \cite{kn:hogy,kn:quench,kn:quenchagain,kn:quenchyetagain} for discussions) that $\hat{q}$ is sensitive to the distinction between central and peripheral collisions, \emph{even in cases where the local energy density is the same} (that is, when one compares central RHIC collisions \cite{kn:RHIC} with sufficiently peripheral LHC collisions \cite{kn:centrality1,kn:centrality2}). This suggests that the local plasma ``knows'' about this distinction through some local effect which tends to reduce $\hat{q}$ in the peripheral case (at a given value of the energy density).

It has been suggested \cite{kn:91} that the enormous \emph{magnetic fields} \cite{kn:skokov,kn:denghuang,kn:review,kn:tuchin,kn:gergely,kn:magnet,kn:hattori} generated in the latter case, but not the former, may have some role to play here; for example, they could influence quenching by affecting momentum diffusion \cite{kn:kitu} or pressure gradients \cite{kn:bali}.
Unfortunately, it is not yet entirely clear whether the magnetic field can survive, with sufficient strength, into the relevant period of the plasma's evolution: this is discussed in, for example, \cite{kn:gursoy,kn:shipu,kn:inghirami,kn:taskforce,kn:arpan,kn:dash,kn:shipu2,kn:shub}. However, the point is that it is at least theoretically possible that magnetic fields might affect jet quenching.

Since vorticity can often have similar effects to magnetic fields (see \cite{kn:volosh,kn:87}, and references therein, for discussions in this specific context), and since it is now clear that vorticity \emph{can} persist throughout the lifetime of the plasma (see \cite{kn:hirono}), it does seem possible that vorticity will affect jet quenching in some way that may be detectable.

However, investigating this from a theoretical point of view is complicated by the following considerations.

One of the peculiarities of the observed polarization effect is that it appears to be more pronounced in collisions at relatively \emph{low} impact energies: see the detailed discussions in \cite{kn:betzgyu,kn:jiang,kn:vortical,kn:lambdabec,kn:hirono}. Thus our attention is drawn to the study of peripheral collisions in the \emph{Beam Energy Scans} \cite{kn:critical,kn:BEStheory} currently under way, and in fact results on jet quenching in such experiments have recently been reported \cite{kn:beamenergyjet}. One can therefore hope that, if indeed vorticity affects jet quenching, this will appear in the data in the reasonably near future, as the BES programmes reach maturity at the RHIC (in the form of the BES II) and in other facilities such as FAIR and NICA \cite{kn:newfair,kn:newbes,kn:newnica,kn:STAR}. However, such investigations will take us into the region of the quark matter phase diagram that is least well-understood theoretically: \emph{the region of large baryonic chemical potential, and of strong coupling} in the plasma regime.

This is the domain in which the holographic or gauge-gravity duality \cite{kn:nat,kn:veron}, based on the AdS/CFT correspondence, has proved to be of service, at least in the qualitative sense of directing our attention to unexpected phenomena such as the low value of the plasma (dynamic) viscosity relative to its entropy density \cite{kn:KSS,kn:SS,kn:bassagain,kn:QGPparameters}; and it is the domain in which duality should be relevant\footnote{This is less clear for the plasmas produced at the LHC, which have such high temperatures that one is less confident that they are indeed strongly coupled.}. We propose to set up a simple gauge-gravity model of the vortical plasma at non-zero values of the baryonic chemical potential\footnote{For recent discussions of holographic approaches to the quark matter phase diagram, see \cite{kn:red1,kn:red2}; for a model incorporating the magnetic field, see \cite{kn:largeBandmu}. For a recent comparison of a holographic prediction regarding the nuclear modification factor \cite{kn:horowitz} with experimental data, see \cite{kn:CMS}. For a holographic account of the evolution of the \emph{shape} of jets, see \cite{kn:jasmine}; for the holography of the effect of anisotropy on jet quenching, see \cite{kn:kazem}. For a non-holographic theoretical approach to possible results on jet quenching from the BES, see \cite{kn:shu}.}, so as to attempt to predict whether we should expect the vorticity to have any detectable influence over the jet quenching parameter, and, if so, in what sense and to what degree.

Thermal systems are represented in holography by a black hole in the dual, asymptotically AdS, bulk. We will need this black hole to be magnetically charged (because we need to be able to distinguish the effects of vorticity from those of the magnetic field), electrically charged (so that we can vary the baryonic chemical potential in the boundary theory independently of the other parameters), and to be endowed with a non-zero angular momentum (to represent vorticity in the boundary theory). Furthermore, we shall choose the topology of the event horizon to be spherical; other topologies are possible \cite{kn:shear}, but they correspond to shearing rather than vortical motion. In short, we need to study dyonic ``topologically spherical'' AdS-Kerr-Newman black holes\footnote{For the use of four-dimensional (uncharged) AdS-Kerr metrics to study other effects of vorticity on the behaviour of the QGP, see \cite{kn:schalm}.}.

It is known \cite{kn:hong1,kn:hong2}, that in a conformal, strongly coupled boundary theory, the quenching parameter $\hat{q}$ scales with the square root of the entropy density $s$. The ratio of $s$ to the energy density $\epsilon$ of the boundary theory has a direct holographic interpretation in a bulk spacetime containing a black hole; so by comparing a suitable AdS-Kerr-Newman black hole with the corresponding charged but non-rotating Reissner-Nordstr\"om black hole, we can study the variation of $\hat{q}$ with vorticity by adjusting the angular momentum of the black hole in the dual theory. (To be precise, this procedure allows us to compute $\hat{q}$ in suitable peripheral collisions, given its value in the corresponding central collisions.)

It turns out that there are some subtleties here. The black hole angular momentum parameter $a$ corresponds holographically to the ratio of the plasma angular momentum density to its energy density, and, as might be expected, the effect of vorticity on the quenching parameter is large when $a$ is large. However, due to a relativistic effect, the collisions producing the highest angular momenta are \emph{not} those producing the highest vorticity, angular momentum and vorticity being related non-linearly in the relativistic case. Thus it may be difficult to associate the highest angular momenta with observed polarization effects, and so it could be difficult to establish that the modifications of the jet quenching parameter, to be discussed here, are indeed due to vorticity.

For this reason, we focus on those collisions involving (according to our model) the highest vorticities, not the highest angular momenta; this is predicted (below) to arise in collisions in the $10 - 15 \%$ centrality bin\footnote{That is, only around $10\%$ to $15\%$ of collisions observed in (for example) the STAR detector system at the RHIC produce more tracks in the Time Projection Chamber than the collisions producing maximal vorticity. Centralities are usually reported in bins, typically $0-5\%, 5-10\%, 10-15\%,$ and so on.}.

We find that the predicted effect of vorticity is always to reduce $\hat{q}$. If the value of $\hat{q}$ at a given impact energy and at almost exactly zero centrality is denoted $\hat{q}^0$, and if the value at maximal vorticity (nominally 10$\%$ centrality) is $\hat{q}^{10}$, then our model predicts that $\hat{q}^{10}/\hat{q}^0\,\approx \, 0.7$. Remarkably, \emph{this same relative reduction in $\hat{q}$ is found at various widely separated points in the the quark matter phase diagram}: at points corresponding to collisions at 200 GeV per pair (where vorticity itself is however difficult to detect), to collisions in the current beam energy scans, at around 19.6 GeV per pair (where vorticity is much more readily detectable), and to collisions in future beam energy scans which will probe the domain of very high baryonic chemical potential \cite{kn:newfair,kn:newnica,kn:STAR}. In every case, the effect of the magnetic field is predicted to be negligible, so any observed effect can confidently be ascribed to vorticity.

Unfortunately, this apparently dramatic reduction may not be readily observable in investigations such as the ones reported in \cite{kn:beamenergyjet}. The most obvious problem is that (current) estimates of $\hat{q}$ are subject to considerable uncertainties of various kinds: for example, \cite{kn:karen} cites (for a quark with energy 10 GeV at time 0.6 fm/c, produced in very central collisions at $\sqrt{s_{NN}} = 200$ GeV) a value of $\hat{q} \approx 1.2 \pm 0.3$ GeV$^2$/fm; clearly our predicted effect will not be readily discernible against background uncertainties of this order.

A further problem is that the maximal effect on jet quenching occurs, as explained above, when the parameter $a$ is maximized, and this occurs in collisions with a surprisingly small centrality (with an impact parameter around 2.5 femtometres), that is, in the same $0 - 5\%$ bin as theoretically perfectly central collisions . One must be concerned, therefore, that reported values of $\hat{q}$ in ``central'' collisions may have in fact been affected by strong vorticity effects in \emph{nearly} central collisions.

We try to circumvent this difficulty by also computing the effect of vorticity on the jet quenching parameter at relatively \emph{large} centralities, around $30 \%$, corresponding to collisions of relatively \emph{low} vorticity (though in fact, in some cases, vorticity is still detectable at such centralities; indeed \cite{kn:STARcoll} focuses on centralities between $20$ and $50 \%$). We find that $\hat{q}^{30}$ is only slightly smaller than $\hat{q}^{0}$; so perhaps the effect we are describing may be more clearly in evidence in a comparison of $10 - 15 \%$ centrality collisions with collisions at large centrality. The problem with this, of course, is that jet quenching itself is not easily measured in this latter case; there are many other effects complicating the analysis of these data, as explained in \cite{kn:beamenergyjet}. Nevertheless, it is reasonable to hope that a reduction in $\hat{q}$ due to vorticity will become apparent in data from FAIR and NICA and in the BES II programme at STAR \cite{kn:newfair,kn:newnica,kn:STAR}, particularly if finer resolution of centrality is possible in studies of jets (as has indeed been achieved in other circumstances, for example see \cite{kn:ATLAS}).

In short, the effect of vorticity is to reduce the jet quenching parameter, \emph{in a characteristic manner}: according to the holographic model, the effect should appear in collisions with centralities around $10 \%$, by comparing the results with those from collisions with either extremely small, or rather large, centralities.

The precise mechanism of this effect, should it exist, is not entirely clear: it may be analogous to the way strong magnetic fields restrict the quark dynamics, affecting the number of microstates and consequently the entropy density \cite{kn:shub}, which in turn is related holographically to the quenching parameter (and also \cite{kn:90} to the kinematic viscosity). This remains to be clarified.

We now proceed to set up the holographic model in detail.

\addtocounter{section}{1}
\section* {\large{\textsf{2. Holography of Vorticity}}}
The dyonic AdS-Kerr-Newman black hole with topologically spherical event horizon has a metric of the form, in Boyer-Lindquist-like coordinates \cite{kn:cognola},
\begin{flalign}\label{A}
g(\m{AdSdyKN)} = &- {\Delta_r \over \rho^2}\Bigg[\,\m{d}t \; - \; {a \over \Xi}\m{sin}^2\theta \,\m{d}\phi\Bigg]^2\;+\;{\rho^2 \over \Delta_r}\m{d}r^2\;+\;{\rho^2 \over \Delta_{\theta}}\m{d}\theta^2 \\ \notag \,\,\,\,&+\;{\m{sin}^2\theta \,\Delta_{\theta} \over \rho^2}\Bigg[a\,\m{d}t \; - \;{r^2\,+\,a^2 \over \Xi}\,\m{d}\phi\Bigg]^2,
\end{flalign}
where
\begin{eqnarray}\label{eq:B}
\rho^2& = & r^2\;+\;a^2\m{cos}^2\theta, \nonumber\\
\Delta_r & = & (r^2+a^2)\Big(1 + {r^2\over L^2}\Big) - 2Mr + {Q^2 + P^2\over 4\pi},\nonumber\\
\Delta_{\theta}& = & 1 - {a^2\over L^2} \, \m{cos}^2\theta, \nonumber\\
\Xi & = & 1 - {a^2\over L^2}.
\end{eqnarray}
Here $- 1/L^2$ is the asymptotic curvature, and $a$ (units of length) is the angular momentum per unit physical mass. We specify ``physical'' here because $M, Q$, and $P$ are \emph{geometric} parameters (with units of length\footnote{Throughout this work we use natural (not Planck) units; it is convenient, since all of the geometric bulk parameters naturally have units of length, to take the base unit to be the femtometre (fm). Recall that $1$ fm$^{-1} \approx 197.3$ MeV.}) which are determined by but are not equal to the corresponding physical parameters: instead \cite{kn:gibperry}, if the physical mass (units of fm$^{-1}$ or MeV) is $m$, the electric charge (dimensionless) is $q$, and the magnetic charge (also dimensionless) is $p$, we have
\begin{equation}\label{C}
m\;=\;M/(\ell_{\mathcal{B}}^2\Xi^2), \;\;\;\;\;q\;=\;Q/(\ell_{\mathcal{B}}\Xi),\;\;\;\;\;p\;=P/(\ell_{\mathcal{B}}\Xi),
\end{equation}
where $\ell_{\mathcal{B}}$ is the gravitational length scale in the bulk. (Recall that this quantity appears in the holographic ``dictionary'' \cite{kn:nat}; in order for holography to be useful it must be small (relative to $L$) so that the bulk theory can be treated as a classical theory.) Notice that we must choose $L$ in such a manner that
\begin{equation}\label{D}
a^2/L^2 \;<\; 1
\end{equation}
is always satisfied; that is, it must be satisfied for the most extreme value of $a$ that we can contemplate. The holographic interpretation of this restriction will be discussed below.

The electromagnetic potential form here is given (see \cite{kn:cognola,kn:87}) by
\begin{flalign}\label{E}
A = &\left(-\,{Qr+aP\,\m{cos}\theta\over 4\pi \ell_{\mathcal{B}} \rho^2}\,+{Q\,r_h+aP\over 4\pi \ell_{\mathcal{B}} \left(r_h^2+a^2\right)}\right)\,\m{d}t \\ \notag \,\,\,\,&+\;\left({1\over 4\pi \ell_{\mathcal{B}}\rho^2}\,\left[Qar\,\m{sin}^2\theta+P\,\m{cos}\theta\left\{r^2+a^2\right\}\right]-{P\over 4\pi \ell_{\mathcal{B}}}\right)\,\m{d}\phi,
\end{flalign}
where $r_h$ is the value of the radial coordinate at the event horizon; the constant terms are determined by requiring that the Euclidean version of this one-form should be well-defined. Notice that, in general, the electric and magnetic fields each depend on both $Q$ and $P$ in the presence of angular momentum\footnote{The reader may find it strange that $Q$ and $P$, rather than $q$ and $p$ (equations (\ref{C})), occur here. To see that this is correct, one can integrate (for example) the electric field, given here by
$$E={-1\over 4\pi \ell_{\mathcal{B}}\rho^4}\left[Q\left(r^2-a^2\m{cos}^2\theta\right)+2Pra\,\m{cos}\,\theta\right],$$
against the element of area in this case, ${r^2+a^2 \over \Xi}\,\m{sin}\,\theta\,\m{d}\theta\,\m{d}\phi$ (for any surface defined by a fixed $r$), obtaining finally $Q/(\ell_{\mathcal{B}}\Xi) = q$, so the physical charge does emerge correctly in Gauss' law.}.

The Hawking temperature of the black hole is given \cite{kn:cognola} by
\begin{equation}\label{F}
T\;=\;{r_h \Big(1\,+\,a^2/L^2\,+\,3r_h^2/L^2\,-\,{a^2\,+\,\{Q^2+P^2\}/4\pi \over r_h^2}\Big)\over 4\pi (a^2\,+\,r_h^2)}.
\end{equation}

The entropy of the black hole is proportional to the area of its event horizon, which takes the unusual form $4\pi (r_h^2+a^2)/\Xi$, so we have
\begin{equation}\label{G}
S\;=\;{\pi\left(r_h^2+a^2\right)\over \ell_{\mathcal{B}}^2\Xi}.
\end{equation}
For holographic purposes it is actually more useful to consider the ratio of the black hole's entropy to its physical mass, given (see the first member of equations (\ref{C})) by
\begin{equation}\label{GG}
{S\over m}\;=\;{\pi\Xi\left(r_h^2+a^2\right)\over M}.
\end{equation}

The ``holographic dictionary'' for this black hole is constructed as follows.

First, note that the conformal boundary of this spacetime can be represented by the metric
\begin{equation}\label{H}
g(\m{AdSdyKN)}_{\infty}\;=\;-\,\m{d}t^2 \;-\;{2a\,\m{sin}^2(\theta)\,\m{d}t \m{d}\phi\over \Xi} \;+\; {L^2 \, \m{d}\theta^2 \over 1 - (a/L)^2\m{cos}^2(\theta)} \;+\; {L^2 \m{sin}^2(\theta)\m{d}\phi^2\over \Xi};
\end{equation}
this is a distinguished representative of the conformal structure, in that $t$ represents proper time for the corresponding observers at infinity.

In this metric, the tangent vectors $\partial_t$ and $\partial_{\phi}$ are evidently not orthogonal, meaning that there is a residual rotational motion (``frame dragging'') even at infinity: it is this peculiar property that permits a holographic formulation of vorticity. The spatial sections have the topology (but not the geometry) of a two-sphere rotating in the $\phi$ direction, and clearly the maximal velocity of rotation is attained at the equator ($\theta = \pi/2$), which has a circumference given by $2\pi L/\sqrt{\Xi}$. This means that the equator is rotating at a relativistic velocity, its circumference being modified in the usual way (\cite{kn:gron}, page 90) by a Lorentz factor $\sqrt{\Xi} = \sqrt{1 - a^2/L^2}$; so the velocity is just $a/L$.

This velocity, assumed constant, is equal to the circumference divided by the period of rotation; if the angular velocity is $\omega$, then the velocity is $\omega L/\sqrt{\Xi}$, and so we have
\begin{equation}\label{HOGWART}
\omega\;=\;{a\over L^2}\,\sqrt{1\,-\,{a^2\over L^2}}
\end{equation}

The plan now is to use this rotational motion to model the vorticity (which corresponds mathematically to twice the angular velocity) of the fluid represented by the field theory\footnote{For the details of the AdS/CFT duality in this case see \cite{kn:sayan}.} at infinity. To see how this works in detail, observe that, for sufficiently small $\theta$, the metric defined by the last two terms in equation (\ref{H}) is indistinguishable from that of an ordinary (round) two-sphere of radius $\hat{L} = L/\sqrt{\Xi}$. Provided that this quantity is sufficiently large relative to the typical size of a vortex in the plasma (which itself occupies a region with an extent of order 5 to 10 femtometres (fm)), we can therefore approximate this sphere by its flat tangent plane at either of the poles. This means that, in order for spurious curvature effects to be suppressed, we must ensure that $a/L$ is close to unity. As this quantity represents the maximal velocity of the system, which is of course close to 1 in natural units, this is quite reasonable: see below for the details.

The flat plane, which as we have seen rotates with an angular velocity given in equation (\ref{HOGWART}), can then be used to represent a vortex in the \emph{reaction plane} familiar in nuclear physics, the plane conventionally described by Cartesian coordinates $z$ (along the directions of the initial momenta of the nuclei) and $x$: putting $\varrho$ = $\hat{L}\theta$ and retaining the angular coordinate $\phi$, we have plane polar coordinates in the reaction plane, corresponding to the $x,$ $z$ coordinates. (The third coordinate, $y$, is parallel to the average angular momentum vector, and also to the magnetic field vector, resulting from a peripheral collision. We focus here exclusively on the physics of the reaction plane, the directions in which vortices are formed and affect the plasma.)

In summary, then, we can set up a holographic model of the plasma vorticity by studying ``holographic vortices'' which inhabit the tangent plane to the sphere at infinity, near to one of its poles. We must however take care to check that, when we propose holographic interpretations of $a$ and $L$, the inequality (\ref{D}) holds, and also that $\hat{L}$ is very much larger than the typical vortex size. These conditions are in fact easily violated, so we have highly non-trivial consistency conditions on the model.

With this setup, we can proceed to the details of the remaining entries in the holographic dictionary.

As always, the Hawking temperature given by (\ref{F}) will be identified with the temperature of the plasma-like matter in the boundary theory, $T_{\infty}$:
\begin{equation}\label{HAWK}
T_{\infty}\;=\;{r_h \Big(1\,+\,a^2/L^2\,+\,3r_h^2/L^2\,-\,{a^2\,+\,\{Q^2+P^2\}/4\pi \over r_h^2}\Big)\over 4\pi (a^2\,+\,r_h^2)}.
\end{equation}

We can identify the entropy per unit mass of the bulk black hole (equation (\ref{GG})) with the ratio of the entropy density $s$ to the energy density $\epsilon$ of the boundary plasma:
\begin{equation}\label{HH}
{s\over \epsilon}\;=\;{\pi\Xi\left(r_h^2+a^2\right)\over M}.
\end{equation}

The parameter $a$ has a natural holographic interpretation as the ratio of the angular momentum density of this plasma to its energy density.

The length scale $L$ will find its interpretation in terms of the velocity parameter $a/L$: as we have seen, this represents the maximal possible rotational velocity in the boundary theory, so it is natural to interpret it as a typical maximal velocity in the vortical plasma.

The next two entries in the ``dictionary'' are obtained by considering the electromagnetic potential (equation (\ref{E})) at infinity:
\begin{flalign}\label{I}
A_{\infty} \,=\, {Q\,r_h+aP\over 4\pi \ell_{\mathcal{B}} \left(r_h^2+a^2\right)}\,\m{d}t\; +\;{P\over 4\pi \ell_{\mathcal{B}}}\left(\m{cos}\theta \,-\,1\right)\m{d}\phi,
\end{flalign}
with corresponding field strength, expressed with respect to a pair of unit (relative to the metric at infinity) one-forms $\hat{\theta}$ and $\hat{\phi}$ parallel to $\m{d}\theta$ and $\m{d}\phi$, given by
\begin{equation}\label{J}
F_{\infty}\,=\,-\,{\Xi \,P\over 4\pi \ell_{\mathcal{B}}L^2}\,\hat{\theta} \,\wedge \hat{\phi}.
\end{equation}
Thus the magnetic field at infinity is now evaluated in the manner familiar from applications of holography to condensed matter \cite{kn:hartkov}: the field is perpendicular to the reaction plane and uniform within it, and its magnitude is given by
\begin{equation}\label{K}
B_{\infty}\,=\,{\Xi \,P\over \ell_{\mathcal{B}}L^2}.
\end{equation}

Next, the holographic expression for the baryonic chemical potential is obtained by taking three times the timelike component of $A_{\infty}$ (equation (\ref{I}) above) and so we have
\begin{equation}\label{L}
\mu_B \,=\,{3\left(Q\,r_h+aP\right)\over 4\pi \ell_{\mathcal{B}}\left(r_h^2+a^2\right)}.
\end{equation}

Finally we need to relate $r_h$ to the other parameters; this is of course done through its definition,
\begin{equation}\label{M}
\Delta_r(r_h)\;=\;(r_h^2+a^2)\Big(1 + {r_h^2\over L^2}\Big) - 2Mr_h + {Q^2 + P^2\over 4\pi}\;=\;0.
\end{equation}

Given physically justified boundary data, we can use the four equations (\ref{HAWK}), (\ref{K}), (\ref{L}), and (\ref{M}), to solve for the four black hole geometric parameters $Q, P, M, r_h$.

Our objective is to study, at suitable points in the quark matter phase diagram, holographic predictions for the behaviour of the jet quenching parameter $\hat{q}$ with respect to variations of the local angular momentum density; this is done by studying the effect of varying the black hole parameter $a$ on the black hole entropy density $s$, since $\hat{q}$ scales with the square root of $s$. To do this, we need to introduce some specific data into the above general formalism.

\addtocounter{section}{1}
\section* {\large{\textsf{3. Numerical Details of the Bulk Geometry }}}
We are concerned here with the effect of vorticity on the jet quenching parameter; holographically, this means that we are interested in the effect of vorticity on the entropy density, computed through equation (\ref{HH}).

As can readily be seen from the above discussion, every parameter on the right side of equation (\ref{HH}) is affected by variations in the vorticity, communicated through the parameter $a$. Although this is far from clear from the form of equation (\ref{HH}) (and one must recall from equation (\ref{HOGWART}) that the relation between $a$ and the vorticity is \emph{not} simple), it turns out (from a numerical investigation) that the strongest effect on the entropy density is felt when the parameter $a$ is maximized.

We begin, therefore, with a discussion of the maximal possible value of $a$, the ratio of the angular momentum density to the energy density. The overall angular momentum acquired by the plasma in the RHIC experiments (at an impact energy of 200 GeV per pair) was estimated in \cite{kn:bec}. It is, for a given model of the nucleus (we use a hard-sphere model, but it is shown in \cite{kn:bec} that a Woods-Saxon model leads to similar results), a definite function of the impact parameter $b$, which is small at small and large values of $b$, attaining a maximum of order\footnote{In natural units, angular momentum is dimensionless; in conventional units it has the same units as action, so ``$7 \times 10^4$'' should be interpreted as ``$7 \times 10^4 \cdot \hbar \,$''.} $7 \times 10^4$ at around (for collisions of gold nuclei, with radii around 7 fm) $b = 2.5$ fm.

A simple computation of the corresponding volume in this ``optimal'' case, regarded as the volume of the intersection of two Lorentz-contracted nuclei of thickness \cite{kn:phobos} $\approx$ 2 fm, leads to an estimated angular momentum density in the vicinity of 400 fm$^{-3}$, so that, with the usual \cite{kn:phobos} estimate of the energy density in early stages of RHIC collisions, around 3 GeV/ fm$^3$, we obtain $a \approx 27$ fm. (Both densities will decline rapidly after this initial stage, but we can assume that their ratio does not change significantly during the interval we are studying here.) However, one sees from \cite{kn:bec} that the angular momentum is sensitive to the impact energy ---$\,$ for example, it is well over an order of magnitude larger, at a given impact parameter, for LHC collisions than for the RHIC collisions we are considering here ---$\,$ so, in some cases, $a$ could easily be considerably larger than this. For the sake of definiteness, we settle on 30 fm $\equiv a_{\m{max}}$ as our maximum value for $a$ for RHIC collisions at 200 GeV per pair; for collisions at lower energies, $a_{\m{max}}$ must be scaled down accordingly.

To determine $L$, we need to estimate a typical maximal velocity of the plasma in the aftermath of a collision at a given impact energy, corresponding here to the quantity $a_{\m{max}}/L$. This is difficult to do, since the forces responsible for converting linear to rotational motion will reduce the velocity from its initial value (which is of course extremely close to the speed of light). Thus $a_{\m{max}}/L$ should be near, but not extremely near, to unity. This is consistent with the fact, mentioned earlier, that the mathematical consistency of the model (which depends on our ability to approximate the sphere at infinity by one of its tangent planes) indeed requires $a_{\m{max}}/L$ to be close to unity.

In view of all this, we shall systematically set $a_{\m{max}}/L = 0.9$; we use this to fix $L$ for all collisions at a given impact energy (so that, for collisions at this impact energy but with ``non-optimal'' impact parameter, $a/L$ is always smaller than $a_{\m{max}}/L$). As we will see, this choice does lead to good agreement with computed values for other parameters, so we have an indirect way of checking that this is reasonable. One can now think of $L$ as having a dual (boundary) interpretation as representing the impact energy. (Before proceeding, note that with this choice we have, with $a_{\m{max}} = 30$ fm as above, $L \approx 33.3$ fm; and the quantity $\hat{L}$ defined earlier is given by $\hat{L} \approx 76.5$ fm, much larger than the region occupied by the plasma, so the boundary sphere is well approximated by its polar tangent plane, as required.)

At this point we pause to consider some important subtleties. It is clear from equation (\ref{HOGWART}) that, if we proceed in this way, the vorticities we compute will be \emph{smaller} at collisions with high impact energies. However, as we have mentioned, this is precisely what theory predicts \cite{kn:betzgyu,kn:jiang,kn:vortical,kn:lambdabec,kn:hirono} and also what is apparently\footnote{However, the statistics are not yet such as to allow a definitive claim that this effect has been observed, and at the present time it is still \emph{possible} that the polarization might be independent of the impact energy.} indicated by the observational data \cite{kn:STARcoll} . Again, this suggests that our procedure here is reasonable.

In this vein, note also that equation (\ref{HOGWART}) shows that the \emph{relativistic} angular velocity is \emph{not} related in a linear way to the angular momentum parameter $a$. For collisions at a given impact energy (so that $L$ is fixed in our scheme), $\omega$ increases with $a$ up to a point, and then begins to decrease: this is required by causality. If we begin with a value of the impact parameter $b$ corresponding to $a_{\m{max}}$, about 2.5 fm here, and then increase $b$, we find that $a$ decreases: but, with $a_{\m{max}}/L = 0.9$, this means that the vorticity (equal here to $2\omega$) actually increases at first, before reaching a maximum of $0.9/a_{\m{max}}$ when $a = a_{\m{max}}/(0.9\,\sqrt{2}) \approx 0.7857\,a_{\m{max}}$, and only then declining. From \cite{kn:bec}, one sees that maximal vorticity occurs for impact parameters around 5 fm, corresponding \cite{kn:bron,kn:olli2} to centralities in the $10 - 15 \%$ centrality bin. (The vorticities reported in \cite{kn:STARcoll} are observed in the $20 - 50 \%$ centrality bins, and so the vorticities considered here are expected to be substantially larger than those values.)

In short, maximal angular momentum corresponds to the maximum effect on the jet quenching parameter, but \emph{not} to maximal vorticity. This is important, because what we observe (in effect) is vorticity, not angular momentum. We will discuss this in more detail, below.

We now turn to $\ell_{\mathcal{B}}$. As mentioned earlier, in order to make use of the gauge-gravity duality, we need to be able to treat the bulk theory classically: that is, we need to ensure that $\ell_{\mathcal{B}}/L$ should be small. We will take $\ell_{\mathcal{B}}/L = 10^{-6}$; fortunately the results are very insensitive to the precise value, once it is this small or smaller\footnote{The necessity of making such a choice is emphasised in \cite{kn:myers}; see also section 2.1.1 of \cite{kn:newmyers} for a related discussion. In principle, this quantity can actually be computed from data, see below. In practice it is more straightforward to postulate a value and show that it leads to agreement with data; again, see below.}.

In order to verify that our values for $a_{\m{max}}$, $L$ and $\ell_{\mathcal{B}}/L$ are reasonable, let us see whether, with these choices, our model can reproduce some  relevant results from the literature. Recently, much attention has been focused on the QGP \emph{equation of state} \cite{kn:bazavov}. Discussions of this have led, in \cite{kn:olli}, to explicit estimates for the entropy density of the QGP at given temperatures, for nearly central collisions. In the case of the RHIC experiments, \cite{kn:olli} reports an expected value of $s = 19$/fm$^3$ for central collisions of gold nuclei at an impact energy of 200 GeV per pair, corresponding to a temperature of around 227 MeV. Assuming an energy density of about 3 GeV/fm$^3$ as above, this means that $s/\epsilon$ is about 1.25 fm.

If we insert $T_\infty = 227$ MeV $= 1.15$ fm$^{-1}$ and (from \cite{kn:phobos}) $\mu_B = 30$ MeV $= 0.15$ fm$^{-1}$, together with our proposed values $L = 33.3$ fm, $\ell_{\mathcal{B}}/L = 10^{-6}$, and $a = B_{\infty} = 0$ (corresponding to central collisions) into equations (\ref{HAWK}), (\ref{K}), (\ref{L}), and (\ref{M}), these can be solved numerically for the four remaining unknowns\footnote{One obtains, in fact, two sets of solutions. This is correct, because the black hole has two horizons, inner and outer, so there must be two values for $r_h$ and therefore for the other parameters. One selects the set corresponding to the outer horizon.} $Q, P, M, r_h$. The latter two can in turn can be substituted into equation (\ref{HH}) to obtain a gauge-gravity prediction for $s/\epsilon$. The result of a straightforward numerical solution of the equations is $s/\epsilon \approx 1.30$ fm, in good agreement with \cite{kn:olli}\footnote{The computed values of $M$ and $r_h$ with these parameter inputs are surprisingly large, $\approx 6.90 \times 10^{7}$ fm and $\approx 5.35 \times 10^3$ fm respectively, showing that the agreement with the results of \cite{kn:olli} is not trivial (that is, not, for example, understandable from dimensional considerations alone).}. (Another way of thinking about this computation is as follows. One could take this value of $s/\epsilon$ for central collisions as an input, and then use the equations to compute $\ell_{\mathcal{B}}/L$, retaining this value for computations in the peripheral case. In practice, this quantity is so small that there are numerical difficulties in solving for it, and it is more straightforward to proceed as we have done here.)

If we consider the corresponding peripheral collisions, the predicted maximal value of the vorticity, computed with our parameter choices using equation (\ref{HOGWART}), is $\approx 0.03$ fm$^{-1}$, just below the value\footnote{This value is computed from the $\sqrt{s_{NN}}$-\emph{averaged} values of the relevant polarizations, and furthermore, as mentioned above, it is deduced from data at $20 - 50\%$ centrality, corresponding \cite{kn:bron,kn:olli2} to collisions at considerably larger impact parameters than those giving rise to maximal vorticity; so the situations here are not directly comparable; however, this value gives a good lower bound on the values of the vorticity that might give rise to polarizations sufficiently large to be observed.} reported in \cite{kn:STARcoll}, roughly $10^{22}\,\cdot\,$s$^{-1} \approx 0.033$ fm$^{-1}$. Thus our model agrees with the fact that polarization of $\Lambda$ and $\bar{\Lambda}$ hyperons is apparently not large enough to be observable at impact energies as high as 200 GeV per pair. (As discussed above, the value of the vorticity corresponding to maximal angular momentum is smaller still: it is $\approx 0.024$ fm$^{-1}$.)

These results suggest that our parameter choices are reasonable.

Having set up a precise bulk geometry in this manner, we now have an explicit holographic model with which to explore the effects of vorticity on the jet quenching parameter.

\addtocounter{section}{1}
\section* {\large{\textsf{4. Variation of $\hat{q}$ with Vorticity at Various Impact Energies }}}
As we have seen, it seems likely that the effects of vorticity in the QGP are most readily detectable not at the maximal RHIC collision energy (200 GeV per pair) but rather at much lower energies \cite{kn:betzgyu,kn:jiang,kn:vortical,kn:lambdabec,kn:hirono}. However, as we will explain, it is interesting to consider the effects of vorticity on jet quenching at several different regions of the quark matter phase diagram.

To do this, we consider theoretical analyses of the Beam Energy Scan (BES) programme conducted at the RHIC facility, and also, more speculatively, consider the situation at facilities under construction such as FAIR. For the BES, the values of the parameters at each collision energy are not known with certainty; for definiteness, we use the recent, clearly set out results given in \cite{kn:sahoo}. (For the magnetic field, we use \cite{kn:denghuang}.) In doing so, we must bear in mind that, at low impact energies and moderate values of the baryonic chemical potential, it is unclear whether a QGP actually forms; according to \cite{kn:sahoo}, the lowest impact energy in the BES for which this can reliably be asserted is 19.6 GeV per pair. Future experiments, such as FAIR, will probe the quark matter phase diagram at very much larger values of the baryonic chemical potential; it is possible that, under those conditions, the QGP can exist at lower temperatures (as the phase transition line is expected to bend downwards in that region of the quark matter phase diagram), so we will consider that case also. In short, we consider $\sqrt{s_{NN}}$ equal to 200, 19.6, and 4.9 GeV, in that order.

\subsubsection*{{\textsf{4.1 Effect of Vorticity on the Jet Quenching Parameter: Collisions at 200 GeV}}}
As we have seen, the most extreme value of the angular momentum occurs in collisions with an impact parameter around 2.5 fm \cite{kn:bec}, leading to an estimate, given above, of $a_{\m{max}} \approx 30$ fm and $L \approx 33.3$. As explained, the value of $a$ in collisions giving rise to the most readily observable vorticities is smaller, around $a = 0.7857\,a_{\m{max}}\,\approx \, 23.6$ fm; this corresponds to $b \approx$ 5 fm. As explained, we do \emph{not} expect the vorticity, even in this most extreme case, to be observable in the polarization experiments when $\sqrt{s_{NN}}$ = 200 GeV, since the predicted polarization (in the notation of \cite{kn:STARcoll}) is only $\overline{\mathcal{P}_{\m{H}}} \approx 1.5 \%$, and indeed this agrees with the observations; but let us proceed to predict the effect on jet quenching, so that we can compare it with the more observationally relevant cases to be discussed later.

The quantity we can actually compute holographically is (see equation \ref{HH}) $s/\epsilon$; assuming a uniform nuclear density (since we do not consider very peripheral collisions, in fact the impact parameter is never larger than 8 fm here, see below), we can use this to study the variation of $s$ with $a$, and then apply the results of \cite{kn:hong1,kn:hong2}.

Denoting the values of the quenching parameter $\hat{q}$ and the entropy density in central (centrality zero) and in ``maximally vortical'' peripheral collisions (nominal centrality $10 \%$) by $\hat{q}^0$, $s^0$ and $\hat{q}^{10}$, $s^{10}$ respectively, we therefore have
\begin{equation}\label{N}
{\hat{q}^{10}\over \hat{q}^0}\;=\;\sqrt{{s^{10}/\epsilon\over s^0/\epsilon}}.
\end{equation}
Our task is to compute $s^{10}/\epsilon$ and $s^{0}/\epsilon$ by inserting the appropriate values (from the discussion above and from \cite{kn:sahoo} and \cite{kn:denghuang}) of $T_\infty, \mu_B, L, \ell_{\mathcal{B}}, a,$ and $B_{\infty}$ into equations (\ref{HAWK}), (\ref{K}), (\ref{L}), and (\ref{M}), solving these four equations for $P, Q, M,$ and $r_h$, and then using equation (\ref{HH}).

For collisions at 200 GeV, taking (from \cite{kn:sahoo}) $T_\infty = 190$ MeV $= 0.964$ fm$^{-1}$ and $\mu_B = 30$ MeV $= 0.15$ fm$^{-1}$ (which agrees with \cite{kn:phobos}), together with (as above) $L = 33.3$ fm, $\ell_{\mathcal{B}}/L = 10^{-6}$, and $a = B_{\infty} = 0$, we can compute $s^0/\epsilon$; substituting these same values for the parameters other than $a$ and $B_{\infty}$, and putting $a = 23.6$ fm and (from \cite{kn:denghuang}, with $b = 5$ fm) $eB \approx 2.3$ m$_{\pi}^2$, where m$_{\pi}$ is the mass of the pion, so $B_{\infty} \approx$ 3.84 fm$^{-2}$, we can compute $s^{10}/\epsilon$. The final result of a straightforward numerical computation is
\begin{equation}\label{O}
{\hat{q}^{10}\over \hat{q}^0}\left(\sqrt{s_{NN}} = 200\; \m{GeV}\right)\;\;\approx\; 0.7062.
\end{equation}
That is, the effect of (maximal) vorticity is to reduce the quenching parameter by about 30$\%$. This puts the effect on the brink of observability with current levels of uncertainty (as mentioned above, \cite{kn:karen} gives $\hat{q} = 1.2 \pm 0.3$ GeV$^2$/fm for central RHIC collisions at early times), leading one to hope that such an effect might be seen in the data from near-future experiments.

We should stress that the value $eB \approx 2.3$ m$_{\pi}^2$ for the magnetic field we have used here is an \emph{upper} bound; it is in fact the estimated \emph{initial} value of the magnetic field. This field can be expected to attenuate to some degree after the collision, though the extent of this attenuation is a matter of debate \cite{kn:gursoy,kn:shipu,kn:inghirami,kn:taskforce,kn:arpan,kn:dash,kn:shipu2,kn:shub}. In any case, we have repeated the above calculation for a wide variety of values of $B_{\infty}$, including zero; in every case, we find that the effect is completely negligible\footnote{Notice that we are dealing here with the situation at the RHIC and in the beam energy scans. It may well be otherwise in the case of LHC collisions, where far larger fields are encountered.}. Thus we can be confident that the effect we are finding here is indeed due to vorticity, and not to the magnetic field.

In practice, there may be difficulties in observing this effect, as was discussed in Section 1. The problem is that observations at ``zero'' centrality may be affected by the fact that the maximal value of the angular momentum parameter $a$ occurs in collisions which are \emph{nearly} central, with impact parameter around 2.5 fm (these also being in the $0 - 5 \%$ centrality bin). It may therefore be preferable to compare the ``maximally vortical'' collisions with collisions at fairly large centrality, in which the vorticity is almost as low as in central collisions (though not so low as to be undetectable). Indeed, if we repeat the above calculations for collisions with impact parameter around 8 fm, corresponding to about $30 \%$ centrality, then we find that $\hat{q}$ is reduced by vorticity (from the value for exactly central collisions) only to a negligible extent (about $3 \%$ instead of $30 \%$). In other words, we have
\begin{equation}\label{P}
{\hat{q}^{10}\over \hat{q}^{30}}\left(\sqrt{s_{NN}} = 200\; \m{GeV}\right)\;\;\approx\; 0.7259,
\end{equation}
which is in practice indistinguishable from $\hat{q}^{10}/\hat{q}^0$. That is, the vorticity-induced drop in the quenching parameter may be most clearly seen if one compares the data at $10 \%$ centrality with the values at $30 \%$.

In both cases, however, the correlation of the effect with vorticity cannot be established, since the hyperon polarization is not detectable in collisions at these energies. Let us therefore turn to lower impact energies.

\subsubsection*{{\textsf{4.2 Effect of Vorticity on the Jet Quenching Parameter: Collisions at 19.6 GeV}}}
According to \cite{kn:sahoo}, collisions at $\sqrt{s_{NN}} = 19.6$ GeV per pair give rise to a plasma with a temperature around 0.867 fm$^{-1}$, a baryonic chemical potential of about 1.015 fm$^{-1}$, and an energy density around $5.57$ fm$^{-4}$; as the peak angular momentum scales with the impact energy \cite{kn:bec}, we find that in this case $a_{\m{max}}$ is about 5.56 fm and $L \approx 6.18$ fm. Maximal vorticity corresponds to $a = 0.7857\,a_{\m{max}}\,\approx \, 4.37$ fm. The magnetic field also roughly scales with impact energy and so $B_{\infty} \approx$ 0.376 fm$^{-2}$. Proceeding in the same way as before, we find
\begin{equation}\label{Q}
{\hat{q}^{10}\over \hat{q}^0}\left(\sqrt{s_{NN}} = 19.6\; \m{GeV}\right)\;\;\approx\; 0.7069.
\end{equation}
Again, if we compare $\hat{q}^{10}$ with $\hat{q}^{30}$ instead of $\hat{q}^{0}$, we find, as before, that this makes little difference:
\begin{equation}\label{R}
{\hat{q}^{10}\over \hat{q}^{30}}\left(\sqrt{s_{NN}} = 19.6\; \m{GeV}\right)\;\;\approx\; 0.7266.
\end{equation}
As in the previous case, the reduction here is indeed due to vorticity: varying the magnetic field has almost no effect.

The similarity of these results to those found above at $\sqrt{s_{NN}} = 200$ GeV is quite remarkable: the more so as all of the individual parameter values differ dramatically in the two cases. For example, in the case of the computation of $s^{10}/\epsilon$ at 200 GeV, one finds that the black hole parameters $M$ and $r_h$ are given by $M \approx 4.064 \times 10^7$ fm and $r_h \approx 4487$ fm, whereas the same parameters arising in the computation of $s^{10}/\epsilon$ at 19.6 GeV are $M \approx 35038$ fm and $r_h \approx 138.7$ fm.

The same difficulties of observation arise in this case as in the previous one; but at least in this case the vorticity should be quite apparent in hyperon polarization experiments, since our model predicts (again in the notation of \cite{kn:STARcoll}) $\overline{\mathcal{P}_{\m{H}}} \approx 9.3 \%$. (This is well above the value reported at this energy in \cite{kn:STARcoll}, but, as mentioned above, this is to be expected, since that reported value corresponds to collisions at much higher centrality than those considered here, producing much lower local angular momentum densities.)

In short, our model predicts a reduction in the jet quenching parameter in collisions at 19.6 GeV, of the same relative magnitude as in the 200 GeV case, at particular values of the centrality, correlated with observably large hyperon polarizations.

We now turn to expectations for future experiments at still smaller values of the impact energy.

\subsubsection*{{\textsf{4.3 Effect of Vorticity on the Jet Quenching Parameter: Collisions at 4.9 GeV}}}
Let us consider the situation at very low impact energies (and correspondingly large values of the baryonic chemical potential), a region of the quark matter phase diagram to be explored in future facilities such as FAIR. Here the input values are of course more speculative: we have been guided by the discussion in \cite{kn:newfair}.

With this understanding, let us consider hypothetical collisions at 4.9 GeV per pair, generating a quark-gluon plasma with a temperature of perhaps 150 MeV or 0.761 fm$^{-1}$, a baryonic chemical potential around 500 MeV or 2.54 fm$^{-1}$, an energy density of perhaps 1.85 fm$^{-4}$, so that $a_{\m{max}} = 4.182$, and a magnetic field on the order of 0.094 fm$^{-2}$. Maximal vorticity corresponds now to $a = 0.7857\,a_{\m{max}}\,\approx \, 3.29$ fm, and computing as before we find
\begin{equation}\label{S}
{\hat{q}^{10}\over \hat{q}^0}\left(\sqrt{s_{NN}} = 4.9\; \m{GeV}\right)\;\;\approx\; 0.7066,
\end{equation}
and
\begin{equation}\label{T}
{\hat{q}^{10}\over \hat{q}^{30}}\left(\sqrt{s_{NN}} = 4.9\; \m{GeV}\right)\;\;\approx\; 0.7263.
\end{equation}
Once again, the magnetic field makes only a negligible contribution to the reduction of $\hat{q}$ seen here.

Again we find that, to three decimal places, the results are identical to those obtained in the previous two cases; again, this occurs despite the fact that the individual parameter values are very different (here $M \approx 7604$ fm and $r_h \approx 68.8$ fm). The agreement is still more remarkable here since our input values are more speculative. Thus the agreement we found earlier can be no mere coincidence, and in fact we have found that these values are universal across the quark matter phase diagram, at least for the physically reasonable domain. We have no explanation of this interesting phenomenon.

The model predicts an extremely high value $\overline{\mathcal{P}_{\m{H}}} \approx 14 \%$ for the polarization in this case; it remains to be seen whether this is at all realistic.

Again, we can summarize by stating that our model predicts a reduction in the jet quenching parameter in collisions at 4.9 GeV, of the same relative magnitude as in the 200 GeV and 19.6 GeV cases, at particular values of the centrality, correlated with very large hyperon polarizations.

\addtocounter{section}{1}
\section* {\large{\textsf{5. Conclusion}}}
It is clear that the holographic model predicts a significant relative reduction in the jet quenching parameter in collisions giving rise to large vorticities: specifically, there should be a reduction in collisions around $10 \%$ centrality relative to collisions at either extremely small or fairly large centrality. This is probably just out of range of analyses of current data, but one can reasonably hope that such an effect will be perceptible in data to be taken in forthcoming experiments, in the relatively near future.

The reduction itself may perhaps be due to the effect of vorticity on the diffusion of momentum in the QGP. This is measured by the kinematic viscosity $\nu$, which is the ratio of the dynamic viscosity $\eta$ to the energy density. This is relevant here, because in a holographic model based on Einstein gravity (only), as is the case here, one does not expect the familiar ratio $\eta/s$ to vary\footnote{It does of course vary, in reality, with respect to variations of temperature and baryonic chemical potential \cite{kn:bassagain,kn:QGPparameters}, but neither of these varies when we compare collisions at different centralities but the same collision energy.} with centrality; but then since we have
\begin{equation}\label{U}
\nu\;=\;{\eta\over \epsilon}\;=\;{\eta\over s}\times{s\over \epsilon},
\end{equation}
this means that $\nu$ is smaller when the vorticity causes $s/\epsilon$ to drop. It would be interesting to see in detail whether this can account for the effect.

More puzzling is the fact that the holographic model predicts that the relative decrease in $\hat{q}$, about $30 \%$ at maximal vorticity, is the same under such radically different circumstances as those found in collisions at $200$ GeV impact energy and their counterparts at far lower energies. If the effect itself is confirmed in near-future experiments, it will be very interesting to see whether holography is correct regarding this, and to investigate possible reasons for it.

\addtocounter{section}{1}
\section*{\large{\textsf{Acknowledgements}}}
The author thanks Dr Soon Wanmei for valuable discussions.

\end{document}